\documentclass[12pt]{article}
\usepackage{graphics}
\input epsf

\newcommand{\be}{\begin{eqnarray}}
\newcommand{\ee}{\end{eqnarray}}

\begin{document}
\title{
\begin{flushright}
{\large UAHEP051}
\end{flushright}
\vskip 1cm
Growth of a Susy Bubble: Inhomogeneity Effects}
\author{L. Clavelli\footnote{lclavell@bama.ua.edu}\\
Department of Physics and Astronomy\\
University of Alabama\\
Tuscaloosa AL 35487\\ }
\date{June 10, 2005}
\maketitle
\begin{abstract}
In a dense star, the Pauli exclusion principle functions as an enormous
energy storage mechanism.  Supersymmetry could provide a way to recapture
this energy.  If there is a transition to an exactly
supersymmetric (susy) phase, the trapped energy can be released with consequences
similar to gamma ray burst observations.  Previous zeroth order calculations
have been based on the behavior in a prototypical white dwarf of solar mass and
earth radius (such as Sirius B) and have neglected density inhomogeneity.
In this article
we show that the effects of density inhomogeneity and of variations in
masses and radii are substantial enough to encourage further exploration of
the susy star model.  In addition, the effects discussed here have
possible applications
to the growth of bubbles in other phase transition models in dense matter.
\end{abstract}
\renewcommand{\theequation}{\thesection.\arabic{equation}}
\renewcommand{\thesection}{\arabic{section}}
\section{\bf Introduction}
\setcounter{equation}{0}

     In the past couple of years there has been a burst of theoretical activity discussing possible transitions between string vacua of differing amounts of supersymmetry \cite{Bousso}.  These include a string theory study of a transition from a local minimum of positive vacuum energy to an exactly supersymmetric phase
\cite{Kachru}.  It is reasonable, therefore, to consider possible
phenomenological signals for such transitions.
We live in a broken-susy phase with, apparently, a positive vacuum energy density
\be
    \epsilon = 3560 {\displaystyle MeV/m}^3
\label{vacenergy}
\ee
leading to an acceleration in the expansion of the universe.  The basic string
theories, on the other hand, suggest a true vacuum of exact supersymmetry with massless
ground state supermultiplets.  In flat space these have zero vacuum energy.
It is, therefore, interesting to consider whether we are living in a false vacuum which will
ultimately decay to an exact susy ground state.  False vacuum decays were treated in some generality
many years ago \cite{Coleman}.  In a false vacuum with energy density $\epsilon$ it is expected
that bubbles of true vacuum with radius $r$ and surface tension $S$ will have an effective
potential given by a sum of a volume term and a surface term,
\be
     V(vac) = - 4 \pi ( r^3 \epsilon /3 - r^2 S)  \quad.
\label{V(vac)}
\ee
For small $\epsilon$ the surface tension $S$ can be treated as a constant.  Bubbles will be
constantly nucleated from the vacuum with a steeply falling distribution in initial radii.
However, only those with initial radii greater than some critical radius
\be
     R_c(vac) = \frac{3 S}{\epsilon}
\label{R_c(vac)}
\ee
will grow to effect a phase transition while smaller bubbles will be rapidly quenched.
The probability per unit time per unit volume to produce a bubble of radius $R_c$ or
greater and, therefore, to effect a phase transition to the true vacuum has been given
\cite{Coleman} in the form
\be
     \frac{d^2P}{dt d^3 r} = A e^{-B(vac)}
\label{AemB}
\ee
where, assuming a thin wall between the phases,
\be
     B(vac) = \frac{27 \pi^2 S^4}{2 \epsilon^3 } \quad.
\label{Bvac}
\ee
In this picture, the fact that our broken susy world has existed so long is due to the
smallness of $\epsilon^3$ relative to $S^4$.  In the physical
vacuum, once a bubble of
critical radius is nucleated the surface will rapidly accelerate and engulf the entire universe.
The fact that this has not happened as yet suggests \cite{Frampton} that
\be
         R_c(vac) > R_{galaxy} \approx 4.7 \cdot 10^{20} m
\label{R_c(vac)estimate}
\ee
or
\be
        S > 5.6 \cdot 10^{23} {\displaystyle MeV/m}^2 = 2 \cdot 10^{-23} M_\circ R_E^{-2} \quad.
\label{Sestimate}
\ee
In much of this article we use the solar mass, $M_\circ = 1.2 \cdot 10^{60}$ MeV, and earth
radius, $R_E = 6.38 \cdot 10^6$ m as
convenient units.  We also use natural units $\hbar=c=1$.

\begin{figure}[htbp]
\begin{center}
\epsfxsize= 4.5in 
\leavevmode
\epsfbox{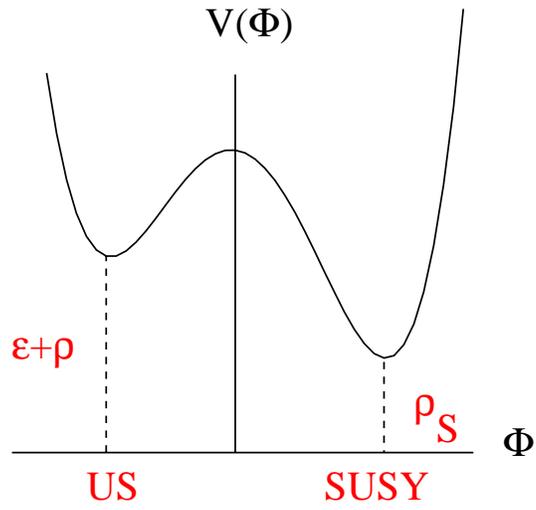}
\end{center}
\caption{The effective potential showing the false vacuum of broken susy and the
true vacuum of exact susy.}
\label{susywell}
\end{figure}

     Reasonable expectations \cite{Gorsky,Voloshin} exist
that vacuum decay will be accelerated in dense matter. Heuristically, this
can be seen by noting that, in the vacuum case,
$\epsilon$ is the energy advantage per unit volume of making a transition to the
exact susy phase.  One might expect, therefore, that the
above equations will be modified in dense matter by replacing $\epsilon$ by the
energy advantage per unit volume of trading the broken susy phase for the exact susy phase,
i.e.
\be
      \epsilon \rightarrow \epsilon + \Delta \rho
\ee
where $\Delta \rho$ is the ground state matter density in the broken susy phase minus the
ground state matter density in the exact susy phase as shown in fig.\  \ref{susywell}.

The difference $\Delta \rho$ is the fermionic
excitation energy density.  The parameter controlling the exponential factor in the
transition rate would then be
\be
      B = \frac{27 \pi^2 S^4}{2 (\epsilon+\Delta\rho)^3 } \quad.
\label{Bmatter}
\ee

The value of $\Delta \rho$ in a white
dwarf star is calculated as follows.  In a degenerate electron gas of $N$ electrons in
a volume $V$ the Fermi momentum is
\be
      p_F = \left ( \frac{3 \pi^2 N}{V} \right) ^{1/3}
\label{Fermimomentum}
\ee
with, assuming equal numbers of neutrons and protons,
\be
      N/V = \frac{\rho}{2 M_N}
\ee
$M_N$ being the nucleon mass.
The average kinetic energy is
\be
      <E> -m = m \left( -1 + {_2F_1}(-1/2, 3/2; 5/2; -p_F^2/m^2) \right) \quad.
\ee
In the limit of zero electron mass which we will use for simplicity, this is
\be
      <E>  = \frac{3 p_F}{4} \quad.
\ee
The kinetic energy density, which is equal to the difference in ground state
energy densities between the broken susy state and the exact susy state, is then
\be
      \Delta \rho = \frac{3 p_F N}{4 V}
            = \frac{1}{4 \pi^2} \left(\frac{3 \pi^2 \rho}{2 M_N}\right)^{4/3} \quad.
\label{delrho}
\ee
We could double this estimate since a comparable contribution is expected from
nuclear excitation energies but, for the present, we will neglect
corrections by factors of order a few.
From \ref{Bmatter} we have
\be
      B = \left( \frac{\tilde\rho}{\rho} \right )^4
\label{B}
\ee
with
\be
       \tilde\rho = \left(\frac{8}{3 \pi^6}\right)^{1/4} S M_N \quad.
\ee

The phase transition probability per unit time per unit volume, $A e^{-B}$,
increases rapidly with
the density of the medium until the density becomes of order $\tilde \rho$
at which point it saturates.  For more dense media, the transition rate is proportional to the
volume.

The longevity of the universe (eq.\ref{Sestimate}) implies that
\be
       \tilde\rho > 0.140 M_\circ R_E^{-3}
\label{rhotildelimit}
\ee
not far from the average density of the prototype white dwarf
\be
       \rho_{WD} = \frac{3}{4\pi} M_\circ R_E^{-3} \quad .
\ee
This suggests that the susy phase transition rate
per unit volume could be appreciable for white dwarf stars but negligible
for less dense objects and for denser objects of much smaller volume.
The relative transition rates in a variety of astrophysical and terrestrial
objects treated as of constant density are tabulated in table \ref{bodies}.
One can see here that if $\tilde \rho$ is close to its lower limit, the
transition rate in white dwarfs is orders of magnitude greater than in
the other considered bodies.  In the future, it would be of interest to explore
larger values of $\tilde \rho$ for which the transition rates for white
dwarf stars and neutron stars are comparable.  However, since the density
distributions in neutron stars and the radiative processes after the
neutron to sneutron pair conversion are greatly different from the
current calculations, these investigations are beyond the scope of the
present paper.

\begin{table}[htbp]
\begin{center}
\begin{tabular}{||lcccc||}\hline
     &     &       &           &                                   \\
     &  M  &   V   &   $\rho$  &   $V e^{-(\tilde \rho/\rho)^4}$   \\
\hline
\hline
     &     &       &           &                                   \\
Cluster core & $10^{14}$   & $4 \cdot 10^{44}$ &   $2 \cdot 10^{-31}$  &  $e^{- 10^{119}}$ \\
Sun          &  $1$        &  $4 \cdot 10^6$    &  $2 \cdot 10^{-7}$    &  $e^{- 10^{23}}$  \\
Earth        & $10^{-6}$   &  $4$              &  $2 \cdot 10^{-7}$    &  $e^{-10^{23}}$  \\
White Dwarf  &    $1$      &  $4$              &  $.2$                  &  $3.7$ \\
Neutron Star &   $1$       &  $0.8 \cdot 10^{-9}$ &  $1.2 \cdot 10^9$      &  $0.8 \cdot 10^{-9}$\\
U$^{238}$ Nucleus& $2 \cdot 10^{-55}$ &  $0.8 \cdot 10^{-63}$ &  $1.2 \cdot 10^{63}$ & $0.8 \cdot 10^{-63}$ \\
     &     &       &           &                                   \\
\hline
\end{tabular}
\end{center}
\caption{
Masses, volumes, mean densities, and relative transition rates for a variety of physical bodies
assuming $\tilde \rho = 0.14$.  Masses and distances are given in units of solar mass and earth
radius.}
\label{bodies}
\end{table}

     In two recent articles \cite{CK,CP} we have explored
the possibility that such a transition in a dense star is the central engine of gamma ray
bursts.  The intense, collimated gamma radiation released in this
way could provide the power to accelerate a macroscopic portion of
the star to relativistic energies as in the cannonball model
\cite{DeRujula}.
For reviews of the susy star idea, see \cite{Pascos04,fnal04}.  In these articles we have
presented zeroth order predictions based on a transition in a typical white dwarf star (e.g. Sirius B)
neglecting density inhomogeneity effects.  In the current paper, we proceed to incorporate these
effects as well as consequences of existing variations in dense star mass and radii.

In dense matter one would expect the critical radius to be
\be
     R_c(r) = \frac{3 S}{\epsilon + \Delta \rho(r)} \approx 12 \pi^2 S \left(\frac{2 M_N}{3 \pi^2
 \rho(r)}\right)^{4/3} \quad.
\ee
In an inhomogeneous medium, a bubble of radius $r$ will grow as long as
\be
     r > R_c(r) \quad.
\label{growthcondition}
\ee
Even ignoring density inhomogeneity within the star, the critical radius could be
quite small inside the star but jump at the surface to a value that is much larger than the radius
of the star.  This has the effect of efficiently confining the susy phase to the interior of the star.

In the exact susy phase the particle and sparticle have a common mass which we have assumed
to be that of the particle in the broken phase.  In order for the phase
transition to proceed, it is a crucial assumption that the common mass is no greater than the particle
mass.  This assumption is, perhaps, supported by the fact that the ground state masses in the exactly
supersymmetric string theories are zero.  Also, susy models with radiative electroweak symmetry
breaking (EWSB) relate the EWSB to the susy breakdown so that, in the absence of susy breaking,
all particle masses vanish.  There are no comparable physical models suggesting the opposite
assumption, namely that the common mass in the exact susy phase is
at higher energy.   Nevertheless, this opposite assumption could also be considered but perhaps
only briefly since, in this case, there is no exothermic phase transition unless, perhaps, the
Fermi energy was greater than the mass difference.

If a SUSY bubble forms in a dense material medium, the interior of the bubble will find itself greatly
out of equilibrium since many of the fermionic constituents occupy high energy levels whereas
scalar particles could all occupy ground state energy levels.  In such a situation
particle pairs will rapidly convert to sparticle pairs.  Electrons, for example,
will pair convert into scalar electrons (selectrons) thus evading the energy storing property of
the Pauli exclusion principle:
\be
       e^{-}e^{-} \rightarrow {\tilde e}^{-}{\tilde e}^{-}
\label{sigma}
\ee
Note that, although the final state is a selectron pair and not a selectron-antiselectron pair, this
process does not require R parity violation.

     Other types of phase transition models have also been proposed to explain gamma ray bursts.  Among these
are transitions to quark matter \cite{Berezhiani}, transitions to a
color superconducting state \cite{Sannino}, and transitions to mirror fermions \cite{Silagadze}.
The calculations of the current paper on the behavior of a susy bubble can also be applied to hypothetical
bubbles of these phases if they begin in the high density regions.

     The susy phase transition model applied to the typical white dwarf star neglecting density inhomogeneity
predicts, correctly though roughly, in a relatively parameter-free way, the minimum duration of
the burst (the light crossing time), the mean gamma ray energy (the mean electron kinetic energy), and the
total burst energy (the total electron kinetic energy).   In addition (and in distinction to the other
phase transition models mentioned above), the fact that the final state of the susy transition consists of
scalar constituents predicts a significant amount of jet collimation due to Bose enhancement.

     In section II of this paper we review the density profile of white dwarf stars produced by the balance
of inward gravitational pressure and outward electron degeneracy
pressure.  The electron momentum distribution differs greatly from
that of a degenerate Fermi gas at uniform density.  In section III
we treat the classical free collapse time of an inhomogeneous susy
star relieved of Pauli blocking, taking into account the full
range of white dwarf stars.  We also compute the variations in
burst duration assuming that, in a dense star, the true vacuum
bubble expands at the density dependent speed of sound which, in
the high density limit approaches $c/\sqrt(3)$. Section IV is
reserved for conclusions.

\section{\bf Density and momentum space gradients in a white dwarf star}
\setcounter{equation}{0}

    Since Chandrasekhar \cite{Chandra}, it has been axiomatic in astrophysics that isolated stars below
a mass of about $1.41$ solar masses are stable against collapse due to electron exchange degeneracy.
The density profile of such white dwarf stars is determined by hydrodynamic equilibrium between
gravity and this outward degeneracy pressure initially augmented by thermal pressure.

\begin{figure}[htbp]
\begin{center}
\epsfxsize= 4.5in 
\leavevmode
\epsfbox{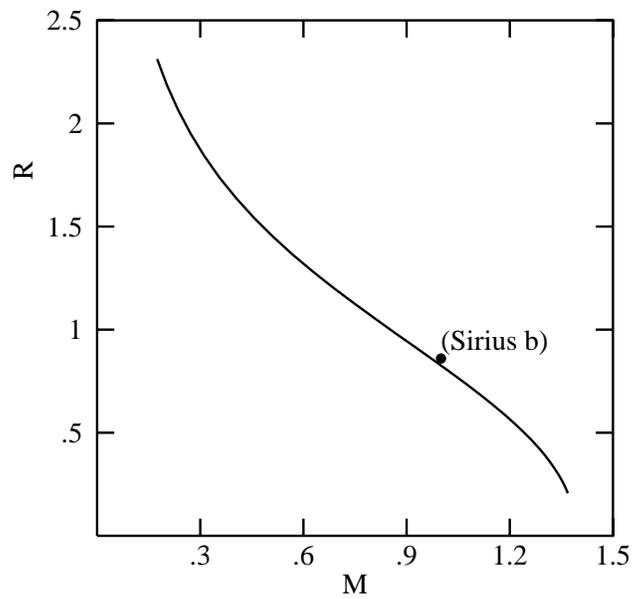}
\end{center}
\caption{The radius versus mass relation for a white dwarf at zero temperature determined by the
balance of gravitational and electron degeneracy pressure.  R and M are measured in units of
earth radius and solar mass respectively.  The prototype white dwarf,
Sirius b, is indicated.}
\label{massrad}
\end{figure}

\begin{figure}[htbp]
\begin{center}
\epsfxsize= 4.5in 
\leavevmode
\epsfbox{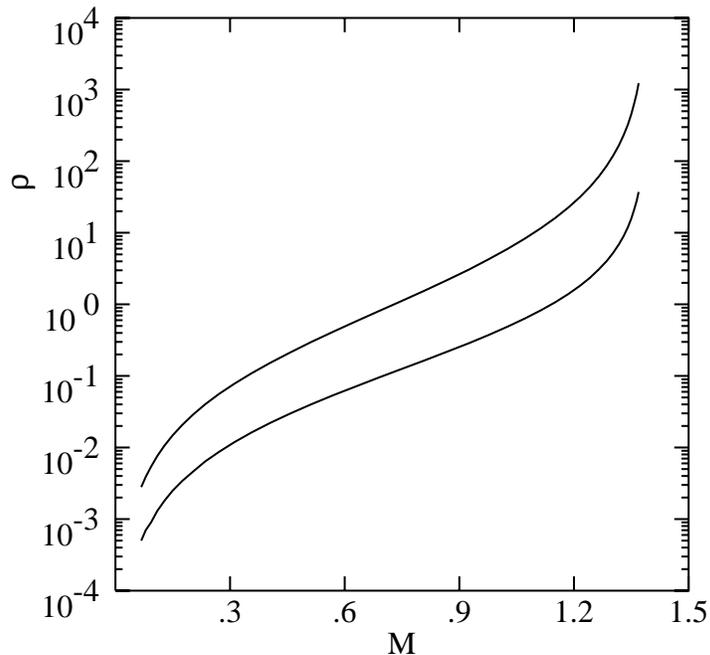}
\end{center}
\caption{The density at the stellar center (upper curve) and the
average density (lower curve) in units of $M_\circ R_E^{-3}$ plotted against the total mass
of the star in units of solar mass $M_\circ$.}
\label{massrho}
\end{figure}

The degeneracy pressure of the electrons in a white dwarf takes a simple form in either the
extreme relativistic or non-relativistic limits.  In intermediate regimes, it can be
written \cite{Harwit} in terms of the variable
\be
     x = \frac{\hbar}{m_e c} \left(\frac{3 \rho(r)}{8 \pi \mu_e M_N}\right)^{1/3} = b \rho(r)^{1/3}
\ee
where $\mu_e=A/Z =2$ and $M_N$ is the nucleon mass.
The degeneracy pressure is proportional to the function
\be
      f(x) = \frac{1}{8} \left ( x(2 x^2 -3)\sqrt{x^2+1} + 3 \sinh^{-1}(x) \right ) \quad;
\ee
Specifically,
\be
      P_d = a f(x)
\ee
with
\be
     a = \frac{8 \pi m_e^4 c^5}{3 \hbar^3} \quad .
\ee
The degeneracy pressure gradient is
\be
     \frac{dP_d}{dr} = \frac{ab}{3} \rho^{-2/3} f'(x) \frac{d\rho}{dr} \quad.
\label{Pdgrad}
\ee
In low temperature equilibrium this must balance the gravitational pressure gradient
\be
      \frac{dP_G}{dr} = - \frac{\rho(r) G_N M(r)}{r^2} \quad.
\label{PGgrad}
\ee
Here, $M(r)$ is the mass within radius $r$ and $G_N$ is the gravitational constant.
The resulting integro-differential equation for $\rho$ can be solved by choosing an arbitrary
starting value $\rho_0$ at the center of the star and integrating outward until the density falls
to zero, recording at each step the value of $M(r)$.  This defines the radius $R$ of the star and
the corresponding mass $M(R)$ as a function of the
peak (central) density.  The resulting mass-radius relation \cite{Chandra}
is shown in fig.\ \ref{massrad}.  In standard astrophysics, all isolated stars with a mass
of less than $1.41 M_\circ$ will ultimately decrease in radius as they cool until they reach a
point on the curve of fig.\  \ref{massrad} at which point they become absolutely stable.
In fig.\  \ref{massrho} we plot the central and average density of white dwarfs at zero temperature
as a function of the stellar mass.  The central densities are more than
an order of magnitude greater than the average densities.

\begin{figure}[htbp]
\begin{center}
\epsfxsize= 4.5in 
\leavevmode
\epsfbox{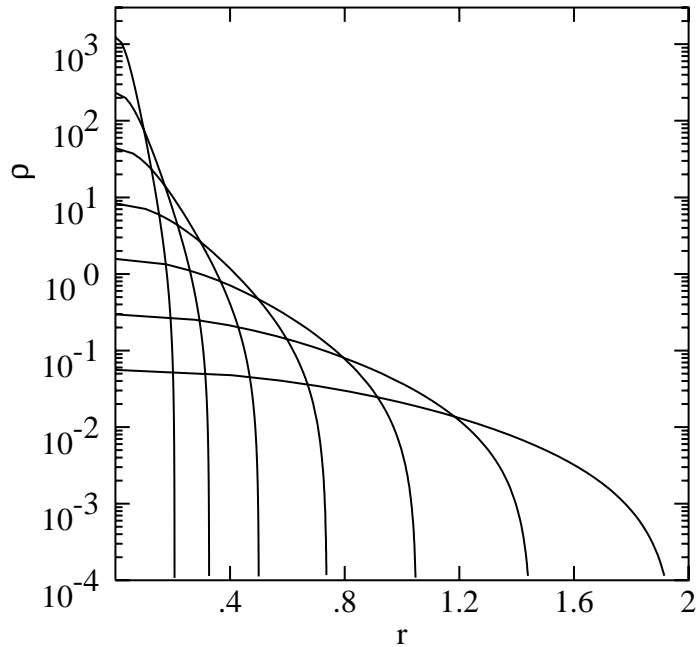}
\end{center}
\caption{Densities as a function of distance from the center of seven representative dwarfs.
The seven masses, in units of the solar mass, are 1.369, 1.330, 1.242, 1.072, 0.811,
0.512, and 0.271.  The dwarfs of higher mass have higher central densities and
greater density gradients.}
\label{rhodist}
\end{figure}

    The density
distribution of these white dwarf stars as a function of distance from the center is a family
of curves of which a representative seven
are illustrated in fig.\  \ref{rhodist}.  Previous work on the susy phase transition in dense stars
\cite{CK,CP} has ignored the strong density variation and relied on average densities only.
The stellar density goes to zero at the surface of the star so, as can be seen from
eq.\ \ref{growthcondition}, the susy bubble will stall at some distance from the surface creating
a thin atmosphere of normal matter.  The thickness of this shell is determined by the surface tension
for which, at present, we know only the limit of eq.\ref{Sestimate}.  The above formulae imply, in this limit,
a skin thickness of about $100 \mu m$ for a star of solar mass and earth radius.  The interior of the bubble constitutes a resonant cavity whose
scalar constituents will continue to radiate gammas until the star radiates all its excitation energy
or collapses under gravitational pressure.

In the inhomogeneous white dwarf, there is a local Fermi momentum given as in \ref{Fermimomentum}
\be
     p_F(r) = \left( \frac{3 \pi^2 \rho(r)}{2 m_N} \right)^{1/3} \quad.
\label{pF(r)}
\ee
The momentum distribution in the electron sea is no longer simply quadratic but takes the step function dependent form
\be
    \frac{dN}{dp} = \frac{4 p^2}{\pi} \int_0^R r^2 dr \theta(p_F(r) - p) \quad .
\ee
The peak electron momentum and therefore the peak photon energy after the phase transition is
given by eq.\ \ref{pF(r)} evaluated at the stellar center, $r=0$.  These peak energies,
tabulated in table \ref{dwarfs} are much
greater than that of white dwarfs treated as of constant density
as in ref. \cite{CK}.  Even higher peak energies, of course,
can be found in stars closer to the Chandrasekhar limit since, there,
the radius approaches zero and the density diverges.  In this region
there is a gradual approach to a neutron star structure.  The gamma ray burst observations,
for comparison, suggest a total burst energy of about $5 \cdot 10^{50}$ ergs and mean photon
energies between $0.1$ and $1.0$ MeV.

In table \ref{dwarfs} we record various properties of
these seven.

\begin{table}[htbp]
\begin{center}
\begin{tabular}{||cccccccc||}\hline
  n  &   M   &   R  &  $\rho_0$  &   $\overline \rho$   &   $p_{max}$   & ${\overline E}$  & $E_{tot}$ \\
\hline
\hline
1 & 0.271 & 1.92  &   0.056 &  0.009 &  0.31 & 0.013 &    $3.5\cdot 10^{48}$\\
2 & 0.512 & 1.44 &    0.297 &  0.041 &  0.54 & 0.034 &    $1.7\cdot 10^{49}$\\
3 & 0.811 & 1.0505 &    1.572 &  0.167 &  0.93 & 0.078 &  $6.1\cdot 10^{49}$\\
4 & 1.072 & 0.7376 &    8.321 &  0.637 &  1.63 & 0.158 &  $1.7\cdot 10^{50}$\\
5 & 1.242 & 0.5002 &   44.053 &  2.366 &  2.84 & 0.291 &  $3.5\cdot 10^{50}$\\
6 & 1.330 & 0.3269 &  233.238 &  9.077 &  4.94 & 0.510 &  $6.6\cdot 10^{50}$\\
7 & 1.369 & 0.2062 & 1234.875 & 37.212 &  8.63 & 0.877 &  $1.2\cdot 10^{51}$\\
\hline
\end{tabular}
\end{center}
\caption{Masses, radii, central densities, average densities, maximum electron momentum , average electron kinetic energy,
and total electron kinetic energy for seven representative white dwarfs at zero temperature. Units of
mass and length are solar mass and earth radius respectively.  The peak
electron momentum is given in MeV/c, the average electron
kinetic energy is given in MeV and the total electron kinetic energy is given in ergs.}
\label{dwarfs}
\end{table}
\section{\bf Factors affecting burst duration}
\setcounter{equation}{0}

     In the susy phase transition model there are several physical effects influencing the
burst duration.  These are
\begin{enumerate}
\item{The bubble growth time.}  This is the time it takes for a bubble nucleated in a high density
region near the center to grow to the stellar surface.  A lower limit to this time is the
light crossing time but, more plausibly, in dense matter the bubble surface should expand
at some rate comparable to the speed of sound in matter of that density.
\item{The light crossing time.} After the bubble has engulfed the star, there could be an
additional time required for light emitted on the far side of the star to cross the stellar
diameter.  At high density, these escaping photons might undergo a random walk leading to a
time proportional to the square of the radius divided by a mean free path.
However, it has been observed \cite{Takahashi} that the Landau-Pomeranchuk-Migdal effect
\cite{Landau}
would greatly increase the transparency of dense matter to gamma rays or, equivalently,
the mean free path.
\item{The free collapse time.}  The conversion of fermions to bosons following a susy phase
transition eliminates the degeneracy pressure and the star will undergo gravitational collapse.
Classically, the star would collapse to a point in a matter of seconds but, once a radius a
few times the Schwarzschild radius is achieved, general relativistic effects begin to
dominate and will cause collapse
to the Schwarzschild radius, as seen by a distant observer, to require an infinite amount
of time during which photon
emission will be progressively red-shifted leading to a certain amount of afterglow below
the gamma ray spectrum.  In ref. \cite{CP} it was found that the
electrons higher in the Pauli tower have a higher probability of
pair conversion implying an earlier pair conversion than that of
lower energy electrons.  This implies a natural progression to smaller frequencies
as the burst progresses.  The growth of the susy bubble into lower density regions
has similar consequences.  These effects by themselves have a much
shorter time scale than observed afterglows and therefore are
relevant only if the gravitational collapse and bubble cooling are
greatly slowed by the following consideration.
\item{Radiation pressure.}  Radiation released by nucleons converting to scalar nucleons,
which then drop into the nuclear ground state,  is expected to rapidly expand the star and,
thus, decrease the
stellar density.  This will increase the bubble growth time, the light crossing time, and
the stellar collapse time.  Density waves could be created in the wake of the initial
blast and might lead to the
rapid time variability observed in gamma ray bursts.  Existing studies
of susy bubble behavior, including the present article, do not include
the effects of radiation pressure.
\end{enumerate}
     We will study first the bubble growth time assuming this is governed approximately
by the speed of sound in dense matter.  The speed of sound
at radius $r$ depends on the local pressure and density and is given by
\be
      v_s(r) = \sqrt{\frac{3P}{\rho}} \quad.
\label{gensoundspeed}
\ee
In the case of constant density this takes the simple form
\be
       v_s(r)  = \sqrt{\frac{2}{3} \gamma \pi G_N \rho (R^2-r^2)} \quad.
\label{soundspeed}
\ee

Here $\gamma$ is the ratio of specific heats ($5/3$ for a monatomic gas), and $R$ is the
stellar radius at which the pressure vanishes.  eq.\ \ref{soundspeed} may be derived by
considering the downward force exerted on a column of matter from radius $r$ to the surface.
The bubble growth time is then
\be
     \tau = \int_0^R dr/v_s(r) = \frac{\pi R}{2 v_s(0)} \quad.
\label{tau0}
\ee

This time is about $2$ s for our typical white dwarf but given the
variations in density among the full sample of white dwarfs, the
growth times based on the average densities as recorded in table\
\ref{dwarfs} have a ratio of maximum to minimum of about $148$.
This is close to the observed ratio of short bursts but far from
the observed ratio of about $10,000$ including long bursts and
short bursts together.  In any case we know from figure
\ref{rhodist} that the densities are rapidly varying especially
for the higher mass dwarfs.  In the case of non-constant density
the pressure at radius $r$ remains governed by the differential
equation of eq.\ \ref{PGgrad}. One begins with zero pressure at
the stellar surface and integrates inward to find the pressure as
a function of distance from the center.  The local speed of sound
is given by eq.\ \ref{gensoundspeed} subject to the limit
$c/\sqrt(3)$ in the high density limit. The bubble growth time is
given by the left-most equality of eq.\ \ref{tau0} . The stellar
mass, radius, and bubble growth time are then given as a function
of the central density. The probability per unit time of a susy
phase transition in a star of radius R is
\be
     \frac{1}{N} \frac{dN}{dt} = 4 \pi A \int_0^R r^2 dr e^{-{(\frac{\tilde \rho}{\rho(r)})^4}} \quad .
\label{prob}
\ee

Because of the exponential suppression at low density, $\tilde \rho$ is a key parameter in
identifying the gamma ray burst progenitor in the phase transition model.  Although $A$ is a
free parameter at this point, the dominant contributions to the bursts are likely to come from
the largest bodies of density $\tilde \rho$ or greater.  The assumption
here is that $A$ is not strongly density dependent.

It is interesting to explore the
possibility that $\tilde \rho$ is close to its lower limit from eq.\ \ref{rhotildelimit}.
This would imply, on the cosmological time scale, an imminent end to our
current world of broken symmetry.  If, on the other hand, $\tilde \rho$ is very much
greater than its lower limit, the bursts would come primarily from denser objects than
most isolated white dwarfs, i.e. either neutron stars or larger bodies in the process of
gravitational collapse.  In this case, density enhancement through accretion such as in the
collapsar model \cite{collapsar}, could play a role.  However, since the Fermi momentum of
the electron sea is proportional to $\rho ^{1/3}$ as in eq.\ \ref{Fermimomentum},  the peak
photon energy potentially gives, in the susy star model, an upper limit on $\tilde \rho$.

\begin{figure}[htbp]
\begin{center}
\epsfxsize= 4.5in 
\leavevmode
\epsfbox{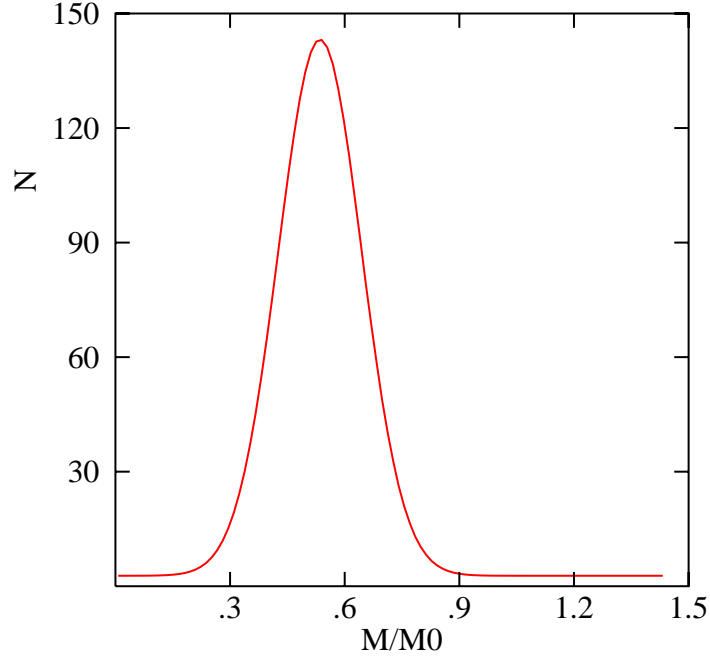}
\end{center}
\caption{Approximate distribution of observed white dwarf masses following ref. \cite{Madej}}
\label{massdist}
\end{figure}

In the current paper we will explore the possibilities that $\tilde \rho$ is related to
its lower limit by a factor of 1,5, or 25.  The rate of bursts as a function of total
progenitor mass is given by multiplying \ref{prob} by the number of white dwarf
stars of given radius or, equivalently, of given mass.  We use the
Sloan survey mass distribution given by \cite{Madej} assuming that the relative numbers
do not change greatly when extrapolated to zero temperature.
The distribution is strongly peaked at $0.56$ solar masses as shown in fig.\ref{massdist}.

The observed burst rate is about
\be
     \frac{dN}{dt} \approx 5 \cdot 10^{-7} {\displaystyle yr}^{-1}{\displaystyle gal}^{-1} \quad .
\ee
but the local rate is much lower suggesting that the fuel of
gamma ray bursts has been largely exhausted and we are now
primarily seeing bursts that happened long ago in more distant
galaxies.  The observed sample of bursts with identified redshifts
seems to cluster around redshift $1$.  This interpretation is,
perhaps, supported by the observed shortage of low luminosity
dwarfs in our galaxy \cite{Winget} and the, at first glance
contradictory, apparent excess of dark objects of white dwarf mass
in the galactic halo \cite{Oppenheimer}.

\begin{table}[htbp]
\begin{center}
\begin{tabular}{||ccccccc|ccc||}\hline
     &       &      &            &                      &             &          &      &      &   \\
  n  &  M     &    R    &    $\tau_0$   &  $\overline\tau$  &  $\tau_c$  &  $E_{tot}$ &  & $\frac{dN}{dt}$ &     \\
     & ($M_\circ$)   &   ($R_E$)  &  (s) & (s) & (s)&(ergs)&  $\tilde\rho=0.14$ & $\tilde\rho=0.7$& $\tilde\rho=3.5$ \\
\hline
\hline
1 & .271  &  1.92  &  1.42  & 12.3  &  7.82 &  $3.5\cdot 10^{48}$ & $1.7\cdot 10^{-11}$ & 0  & 0  \\
2 & .512  &  1.44  &  0.693  &  5.80  &  3.70  &  $1.7\cdot 10^{49}$ & $\underline{18}$    & $1.9\cdot 10^-7$  & $4.4\cdot 10^{-44}$  \\
3 &  .811 &  1.05  &0.348  & 2.86  &  1.82 &  $6.1\cdot 10^{49}$ & $0.78$  & $12$ & $7.9\cdot 10^{-8}$  \\
4 &   1.07 & .737  & 0.181  & 1.47  &  0.93 &  $1.7\cdot 10^{50}$  & $0.46$  & $\underline{20}$  & $31$ \\
5 &   1.24 &  .500  & 0.095  & 0.76  &  0.49 &  $3.5\cdot 10^{50}$  & $0.19$  & $11$  & $\underline{44}$ \\
6 &   1.33 &  .327  & 0.049 & 0.39  &  0.25 &  $6.6\cdot 10^{50}$  & $0.062$ & $4.4$ & $24$ \\
7 &   1.37 &  .206  & 0.024 & 0.19  &  0.12 &  $1.2\cdot 10^{51}$  & $0.017$ & $1.4$ & $9.2$\\
\hline
\end{tabular}
\end{center}
\caption{relative burst rates with given bubble growth times,
constant density growth times, and classical collapse times, as well as
total burst energies.  The burst rates are given with separate arbitrary normalizations
for three values of $\tilde \rho$.  The peak of the
probability distribution is underlined.}
\label{transrates}
\end{table}

From the total energy release, table\ \ref{transrates}
might suggest a $\tilde \rho$ value near or slightly above $3.5 M_\circ R_E^{-3}$.
However, the investigation of beaming and other contributions to the
energy release is at too early a stage to rule out larger values of
$\tilde \rho$.  
Since the bubble growth time is a monotonic function of dwarf mass, we can use the mass distribution
of fig.\  \ref{massdist} and the transition probability of eq.\ \ref{prob} to calculate the shape
of the growth time distribution.  This is shown in table. \ref{transrates} for the three chosen values of
$\tilde \rho$.  It is encouraging that the growth times are in the range of the observed short burst
durations but a critical test of the model must await the incorporation of the other factors
influencing the burst duration as enumerated above.

One of these factors is the collapse time of a star relieved of Pauli blocking.  In \cite{CK}
we have noted that the classical collapse time of a susy white dwarf is
\be
    \tau_c = \frac{\pi}{2} \left( \frac{8 \pi G_N \rho}{3} \right)^{-1/2} \quad .
\ee
Although that paper considered only stars of constant density, the collapse time remains
the same for inhomogeneous stars as long as one uses the average density.  The classical
collapse times of the representative sample of white dwarfs is given in table \ref{transrates}.  However
once the star approaches the Schwarzschild radius, general relativistic effects dilate the
collapse time as seen by a distant observer.  The approach to the Schwarzschild radius,
$r^* = \frac{2 G_N M}{c^2}$ is given by
\be
      r - r^* = (r_0 - r^*) e^{-(t-t_0)c/r^*} \quad .
\ee
During this time the star can still radiate although the photon energies will be redshifted
from their emission energies according to the relation
\be
     E = E_{em} \frac{e^{-(t-t_0)c/{2 r^*}}}{\sqrt(1 + e^{-(t-t_0)c/r^*})} \quad .
\ee

The time dependence of this component of the afterglow is
independent of frequency.
The time constant, $c/r^*$ for this contribution to the afterglow will be of order of ten microseconds
for a star of near solar mass.  With currently available techniques, it will be impossible to observe such a
short afterglow.  Another source of afterglow will be the emission from circumstellar material
irradiated by the burst.  This second source may be absent if the burst comes from the decay
of an isolated star.  Since, at present, afterglows have only been definitely observed for some of the long bursts,
it might be interesting to consider a proposal where all of the short bursts and many of the long bursts originate in isolated stars
while the bursts with extensive afterglows originate in stars with
significant circumstellar material as in the collapsar model.
A very massive star in the process of gravitational collapse will
necessarily pass through stages of high fermion degeneracy where
the possibility of a susy phase transition might become significant
even though the time spent in these stages is not long.

\section{\bf Conclusions}
\setcounter{equation}{0}

We have considered the effects of mass variations and density inhomogeneities in white dwarf
stars under the assumption that such stars experience a phase transition to the exact susy
ground state.  It is clear that much work remains to be done.  Nevertheless, it is
encouraging that the basic assumption with few free parameters produces a gamma ray
emission on the right time scale, with mean gamma energy and total energy release close to
observations.  The density inhomogeneities increase the expected peak photon energy by an
order of magnitude over that of constant density stars of the same average density.

The assumption that the surface tension of a susy bubble is independent of density needs to
be examined and probably relaxed.  Similarly, the strong density dependence of eqs. \ref{AemB},\ref{Bmatter}
might be
somewhat softened by corrections to the thin wall approximation of the vacuum decay studies
\cite{Coleman}.
Effects due to the radiation pressure at non-zero temperature
need to be incorporated and may significantly affect the duration distribution.  Radiation from
the collapse of the Pauli tower in nuclei as well as the enhanced energy release from snuclear
fusion and beta decay need to be studied.  The resulting radiation pressure is expected to
slow the final collapse of a star relieved of Pauli blocking.
In addition, we hope to explore the possibility that the profound nuclear
explosion at the stellar center due to the phase transition sets up density standing waves.  This
could be related to the spikey behavior of the observed gamma ray bursts.

{\bf Acknowledgements}

    This work was supported in part by the US Department of Energy under grant DE-FG02-96ER-40967.
We gratefully acknowledge discussions with Paul Cox, Ben Harms, Irina Perevalova,
George Karatheodoris, Phil Hardee, Bill Keel, and Sanjoy Sarkar.

\end{document}